\documentclass[review]{elsarticle}
\usepackage{lineno,hyperref}
\usepackage[version=3]{mhchem} % Formula subscripts using \ce{}
\usepackage[utf8]{inputenc}
\usepackage[T1]{fontenc}
\usepackage{gensymb}
\usepackage[a4paper, total={6in, 8in}]{geometry}
\usepackage{indentfirst}
\begin{document}

\begin{frontmatter}
\title{Atomistic simulations of surface reactions in ultra-high-temperature ceramics: O$_2$, H$_2$O and CO 
adsorption and dissociation on ZrB$_2$ (0001) surfaces}
\author{Yanhui Zhang$^{a,b}$}
%\ead{yhzhang@ysu.edu.cn}
\address[mymainaddress]{School of Materials Science and
Engineering, Yanshan University, Qinhuangdao, Hebei 066004, China}


\author{Stefano Sanvito$^b$}
\ead{sanvitos@tcd.ie}
\address[mymainaddress]{School of Physics and CRANN, Trinity College Dublin, Dublin 2, Dublin, Ireland}

\begin{abstract}
Understanding surface reactivity is crucial in many fields, going from heterogeneous catalysis to materials
oxidation and corrosion. In order to better decipher the initial stage of surface reactions of ZrB$_2$ exposed 
to the harsh environment of aerospace components, the chemical activity of both Zr- and B-terminated (0001) 
surfaces is predicted and compared by using state-of-the-art density functional theory. In particular the
adsorption, dissociation and diffusion of O$_2$, CO and H$_2$O are extensively examined through the calculation
of the surface adsorption energies and reaction pathways. We find the dissociative adsorption of O$_2$
dominating the reactivity of both Zr- and B-surfaces, while the dissociation of H$_2$O and CO is weakly active
on Zr-terminated surfaces, and even less activated on B-terminated ones. Importantly, we discover that the
reaction of O$_2$ and H$_2$O can trigger surface reconstruction at the B termination, an efficient mechanism for
B removal. Our work thus provides thermodynamic and kinetic insights into the elementary reactions of the most
dominant gases found in the environment typical of aerospace applications, and highlights the diverse surface
reaction mechanisms when ZrB$_2$ exposed to O$_2$, CO and H$_2$O. 
\end{abstract}

\end{frontmatter}


%\begin{tocentry}
%
%Inside the \texttt{tocentry} environment, the font used is Helvetica 8\,pt, as required by \emph{Journal of the American Chemical Society}.
%
%The surrounding frame is 9\,cm by 3.5\,cm, which is the maximum permitted for  \emph{Journal of the American Chemical Society} graphical table of content entries. The box will not resize if the content is too big: instead it will overflow the edge of the box.
%
%\end{tocentry}

\section{Introduction}
Ultra-high-temperature ceramics (UHTCs) form a family of materials defined by an extreme temperature range
of operation, which includes transition metal carbides, nitrides and diborides. UHTCs are very actively
researched for extensive application in various aerospace fields (as thermal protection systems in hypersonic or
atmospheric reusable re-entry vehicles, propulsion components, combustion chambers, engine intakes and rocket
nozzles), in energy (in nuclear fission and fusion, energy harvesting, concentrated solar power), in materials
processing (as high-temperature electrodes, high-speed machining tools, molten metal containment), in
microelectronics (as conductors, barrier layers, lattice-matched substrates for semiconductors), just to name a
few.~\cite{Padture2016,Zeng2017,Fahrenholtz2017,Cahill2019} Typically, UHTCs are characterised by the 
exceptional combination of mechanical, thermal and electrical properties, coupled with melting temperatures in
excess of 3000~$\degree$C. These features allow UHTCs to work in extreme environments as high-temperature and
aggressive atmosphere. However, the attack of the chemical agents usually found in real-life hybrid-fuel 
rockets, namely O$_2$, H$_2$O and CO, still pose huge challenges to the structural integrity of many
components. In fact, these induce changes in the surface chemistry, causing corrosion and facilitating 
rapid ageing. For this reason, a comprehensive understanding of the UHTCs surface reactivity is essential for
improving their structural stability and thus ultimately the safety of UHTC-based parts. 
%To improve the resistance to chemically aggressive environment, researchers have been working on two potential strategies.
%One is to develop the coatings system called environmental-barrier coatings (EBCs). But it unavoidably brings up the issues
%of complex interfaces, which are formed within EBCs or between EBCs and ceramic matrice. The related studies are still
%active to find one EBC system meets all the operating-temperature goals adequately.~\cite{Padture2019} An alternative way
%is to develop UHTC with resistance to surface-reactions. The qualified material surface should react with chemical agents
%meanwhile generating self-protecting films to stand further environmental erosion. One example is the
%Zr$_{0.8}$Ti$_{0.2}$C$_{0.74}$B$_{0.26}$ with superior ablation resistance at temperatures from 2000 to 3000 $\degree$C 
%thanks to the sealing ability of its oxides with suitable viscosity and low vaporization rate.~\cite{Zeng2017} 
%The formation mechanism of different films under various atmosphere carries important information for the tuning of
%corrosion resistance. To tackle this problem, extensive experimental and theoretical studies have been carried out to 
%understand the surface catalysis process, like the adsorption of chemicals, the dominate reactions and related barriers. 

Several experimental investigations have been conducted to unveil the primary surface reactions of various
UHTCs, although most of these studies are only focused on adsorption intermediates or dissociation products. For
instance, dissociative adsorption of O$_2$ was observed on ZrB$_2$(0001) by using high-resolution electron
energy loss spectroscopy (HREELS)~\cite{Aizawa2002a}. Belyansky and Trenary~\cite{BELYANSKY1999191} have
scrutinized the surface activity of the HfB$_2$(0001) and Hf(0001) surfaces, with the help of X-ray
photo-emission spectroscopy (XPS) and RAIRS (reflection adsorption infrared spectroscopy). Molecular adsorption
of CO was found on HfB$_2$(0001), while its dissociative adsorption was observed only on Hf(0001). In addition,
the adsorption of O on the 3-fold hollow site of HfB$_2$(0001) was measured by an impact-collision ion
scattering spectroscopy (ICISS) study~\cite{HAYAMI1994}. Similar studies have been carried out for carbides as
well. For instance, the oxidation process of ZrC was investigated with angle resolved photo-emission
spectroscopy (ARPES)~\cite{Kato_2000}, which identified carbon depletion with the concomitant Zr partial
oxidation in the form of ZrO$_x$ (1 $\leq$ $x$ $\leq$ 2). Furthermore, at an oxygen-modified ZrC(100) surface,
the dissociation of H$_2$O into hydroxyl (OH) and atomic O species was discovered by a combination of
ultraviolet photoelectron spectroscopy (UPS) and XPS measurements.~\cite{KITAOKA200123}

As an alternative to experiments, atomistic simulations provide a powerful tool to explore surface reactions in 
more details. For example, by using density functional theory (DFT) calculations, Say\'{o}s et al.
~\cite{Gamallo2013} proposed that the adsorption of atomic O on the ZrB$_2$ (0001) surface takes place
predominantly at the hollow and B-B bridge sites. Zhou et al.~\cite{Sun2019,Sun2016} emphasized the effects of
the valance electron concentration on the surface stability and oxygen adsorption. The theoretical study of
oxidation of ZrB$_2$ (0001) surface from Cheng et al~\cite{Cheng2018} suggested that the initial oxygen coverage
has only little effect on the oxidation reaction. The similar feature was also claimed for H$_2$O adsorption on
TiO$_2$~\cite{Valdes2019}. Cristol et al.~\cite{Osei-Agyemang2014a} performed a statistical thermodynamics study
of O$_2$, H$_2$ and H$_2$O on a bare and ZrO-modified ZrC (100) surface. They suggested that the oxygen
modification activates the (100) surface, and water adsorbs strongly either as atomic O with H$_2$ release or
into surface OH and H groups. Despite these early studies, a systematic investigation of the surface activity of
UHTCs, looking on an equal footing at O$_2$, CO and H$_2$O, is still missing. This is not ideal, since the
reaction kinetic and its atomic characteristics are both required to effectively guide the tailoring of UHTC
materials against oxidation and corrosion.  

Among the UHTC family ZrB$_2$ stands out because of its excellent chemical stability, high hardness and
strength, great thermal-stress resistance and good electrical and thermal conductivities.
~\cite{Fahrenholtz2007}. In this work, we study in detail the surface activity of ZrB$_2$ with the most common
chemical agents: O$_2$, H$_2$O and CO. These are among the most abundant chemical agents in the combustion gas
of hybrid-fuel rockets. The paper is organised as follows. A brief review of the atomic models (surface slabs,
adsorption sites and adsorbate geometries) and the calculation details are provided in section 2. Then, an
examination of atomic adsorption of O, H, OH and C on Zr-terminated (0001)$_\mathrm{Zr}$ and B-terminated
(0001)$_\mathrm{B}$ surfaces, and the related surface distortions are introduced in section 3. The adsorption
and dissociation mechanism of O$_2$, H$_2$O and CO are separately presented in sections 4, 5 and 6, while a
comparison of the surface reactivity for the three molecules is discussed in section 7. Finally section 8
contains the main concluding remarks. Supplementary Information (SI) complements the materials included here.

\section{Methodology}
\subsection{Surface models}
In order to investigate the ZrB$_2$ reactivity we construct 2$\times$2$\times$7 (0001) surface slabs, whose 
structure is reported in Fig.~\ref{fig1}(a) and \ref{fig1}(b). In particular, we consider both Zr [see
Fig.~\ref{fig1}(a)] and B [see Fig.~\ref{fig1}(b)] terminations, which are denoted respectively as
(0001)$_\mathrm{Zr}$ and (0001)$_\mathrm{B}$. These are selected, since they can be easily stabilized in either
a Zr- or a B-rich chemical environment, as validated both in experiments and theory 
~\cite{0953-8984-20-26-265006,Suehara2010,Zhang2018}. Our surface slab models are deliberately built up with a
symmetric termination so as to reduce spurious dipole effects. In general, periodic boundary conditions are
applied in the $x$-$y$ plane, while a vacuum region of around 26~\AA\ is added along the $z$ direction 
(perpendicular to the surface) to avoid the artificial surface-surface interaction between the periodic
replicas. 

\subsection{Adsorption sites}
The inequivalent adsorption sites over the (0001)$_\mathrm{Zr}$ and (0001)$_\mathrm{B}$ surfaces are identified 
within the smallest structural unit of each surface, namely the threefold Zr-triangle and the six-member B ring 
depicted by the dotted lines in Figs.~\ref{fig1}(a) and \ref{fig1}(b), respectively. For (0001)$_\mathrm{Zr}$
there are three  inequivalent adsorption sites, an hollow site (\textbf{H1}), a bond bridge one (\textbf{B1})
and the atop (\textbf{A1}). Similarly, four sites are identified at the (0001)$_\mathrm{B}$ surface, two hollow
positions (\textbf{H2} and \textbf{H3}), one bond-bridge site (\textbf{B2}) and one atop (\textbf{A2}). An
additional possible adsorption site, denoted as \textbf{B$^{\ast}$} in Fig.~\ref{fig1}(b), always relaxes to the
\textbf{H2} position so that is deemed unstable and it is excluded from further study. 

\subsection{Adsorbate geometries}
\begin{figure}[h!]
\centering
\includegraphics[scale=0.85]{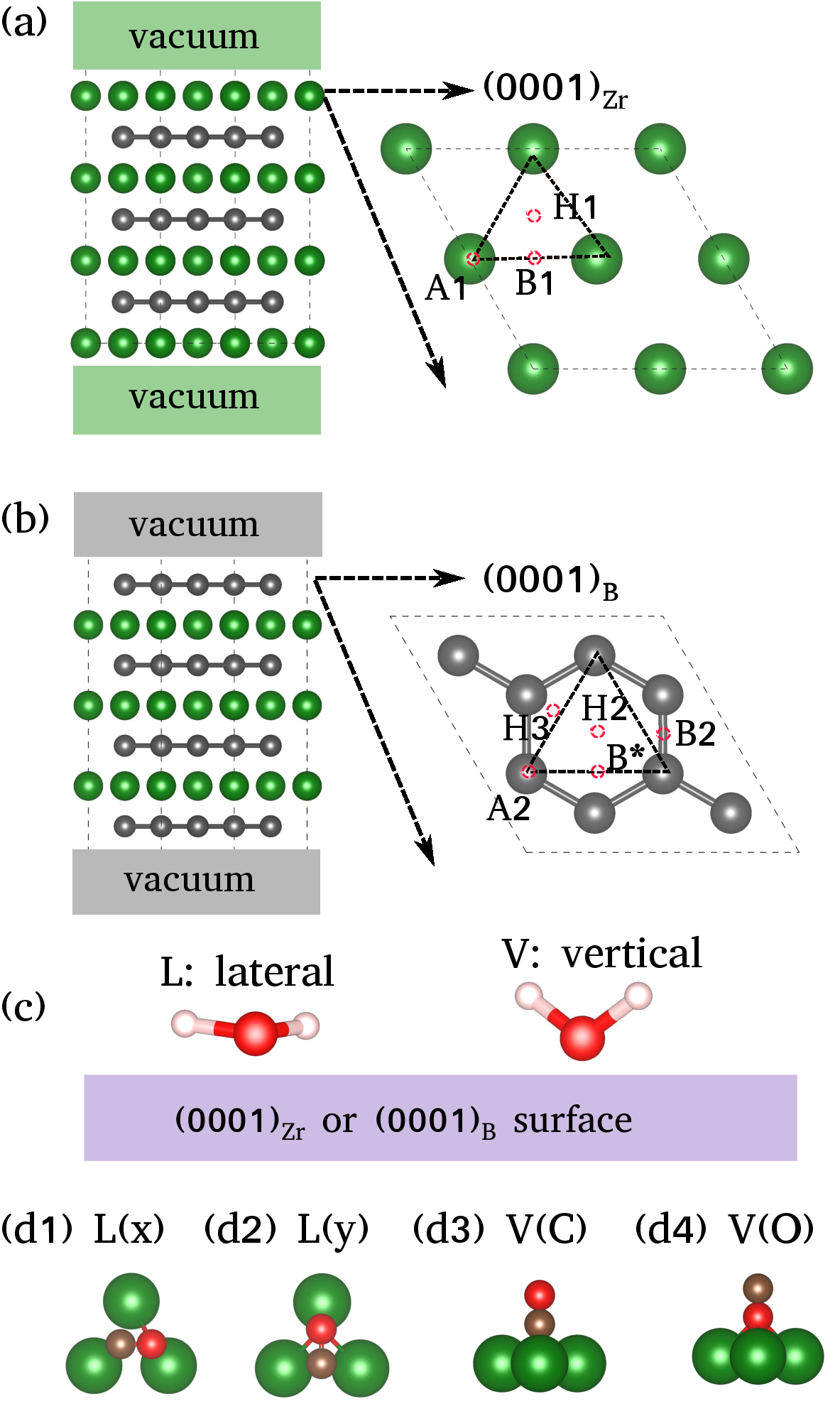}
\caption{Surface slab models of (a) (0001)$_\mathrm{Zr}$ and (b) (0001)$_\mathrm{B}$ slab with side (left-hand
side panels) and top (right-hand side panels) views. The inequivalent adsorption sites (hollow: \textbf{H};
bond-bridge: \textbf{B}; atop: \textbf{A}) are marked by red circles on the top views. Note that two different
hollow sites (\textbf{H2} and \textbf{H3}) are found on (0001)$_\mathrm{B}$, while the \textbf{B}$^{\ast}$
position is unstable and relaxes into \textbf{H2}. The adsorbate geometries are labelled according to the
adsorption sites, the approaching ways and the near-end species and/or alignment axis. The approaching geometry
can be either lateral, \textbf{L}, or vertical, \textbf{V}. These two options are illustrated for H$_2$O 
in panel (c), and further for CO adsorbing on the hollow site of (0001)$_\mathrm{Zr}$ in (d1) \textbf{H1\_L(x)},
(d2) \textbf{H1\_L(y)}, (d3) \textbf{H1\_V(C)} and (d4)  \textbf{H1\_V(O)}.  Colour code: Zr green; B grey; C
brown; O red; H pink.}
\label{fig1}
\end{figure}

Besides the inequivalent adsorption sites the adsorbates can also be distinguished from each other by the way
they approach the surface, namely by their adsorption geometry. Broadly speaking one can separate between the
situations in which all the molecule atoms lay parallelly with the surface plane (lateral geometry, \textbf{L}),
or when they are perpendicular to the surface (vertical geometry, \textbf{V}). These two situations are
illustrated for H$_2$O in Fig.~\ref{fig1}(c). Furthermore, for hetero-molecules like CO the adsorption geometry
can be further characterised by the alignment of the molecule with respect to the crystalline axes for the
\textbf{L} case and by the specific atom closest to the surface in the \textbf{V} situation. Thus, for CO we
have two inequivalent lateral orientations, denoted as \textbf{L($x$)} and \textbf{L($y$)} in
Figs.~\ref{fig1}(d1) and \ref{fig1}(d2), respectively (for the Zr-terminated surface), and two inequivalent
vertical ones, \textbf{V(C)} and \textbf{V(O)} [see Figs.~\ref{fig1}(d3) and \ref{fig1}(d4)]. The different
adsorption states \textbf{H1\_L(x)}, \textbf{H1\_L(y)}, \textbf{H1\_V(C)} and \textbf{H1\_V(O)} will be
examined in detail in section 6.  
The adsorption energy is then simply calculated as,
\begin{align}\label{Etot}
   \Delta E_\mathrm{ads}=E^\mathrm{sub}+E^{x}-E^{x/\mathrm{sub}} \:,
\end{align}
where, $E^{x/\mathrm{sub}}$, $E^\mathrm{sub}$ and $E^{x}$ are, respectively, the DFT total energies of the
surface slab including the $x$ adsorbent, the clean surface and the adsorbent either in gas phase or as an
isolated atom. The value of $\Delta E _\mathrm{ads}$ (eV/atom) measures the reactivity of the various surface
sites and the binding strength of the various adsorbate-adsorbent pairs. 

\subsection{DFT calculation details}
All the calculations are performed using the plane-wave basis-projector augmented-wave method
(PAW)~\cite{Blochl1994} as implemented in the VASP~\cite{Kresse1999} code. The generalized gradient 
approximation (GGA) with the Perdew-Burke-Ernzerhof (PBE)~\cite{Perdew1996} parametrization is adopted to
evaluate the exchange and correlation energy. Recently, it has been shown that, in general, dispersion
interactions strongly influence adsorption geometries and binding energies~\cite{Saqlain2018}.
Here we include these by mean of the DFT-D2 correction~\cite{Grimme.2006}, which has been proved adequate
in the description of transition-metal diborides~\cite{Zhang2018}. The calculated adsorption energies are
compared to those obtained with rPBE+D2 and rPBE+D3 in Fig.~S1 of the SI, showing little variations with the
details of the PBE functional and the van der Waals corrections chosen. For all surface calculations the
Brillouin zone is sampled by using the Monkhorst-Pack method with a $k$-mesh of 8$\times$8$\times$1. The
plane-wave kinetic energy cutoff is set to 500~eV. Geometry optimization is performed until the energy and the
forces are converged within 10$^{-7}$~eV and 0.01~eV/\AA, respectively, and dipole corrections~\cite{Makov1995}
are applied [see the electrostatic potential profile after dipole correction in Fig.~S2 of the SI]. 
Lastly, note that spin-polarization must be included when computing the total energy of the isolated species O,
C, OH, O$_2$ and H, while it is not relevant for the slab calculations.

The minimum energy paths for the molecules dissociation are refined by using the nudged-elastic band (NEB)
method~\cite{Henkelman2000,Henkelman2000a}. Thereafter, the transition states (TS) are located by using the
climbing-image NEB algorithm with a force threshold of 0.01~eV/\AA. During the relaxation of each reaction path,
the adsorbent atoms on the first layer of the slab are allowed to move away from their high-energy and
high-stress positions. This scheme enables us to probe the surface dissociation process, to assess whether there
are kinetic bottlenecks in the surface reaction, and to monitor the accompanying distortions of the adsorbents. 

\section{Atomic adsorption}\label{AtomicAds}
\begin{figure}[h!]
\centering
\includegraphics[scale=0.25]{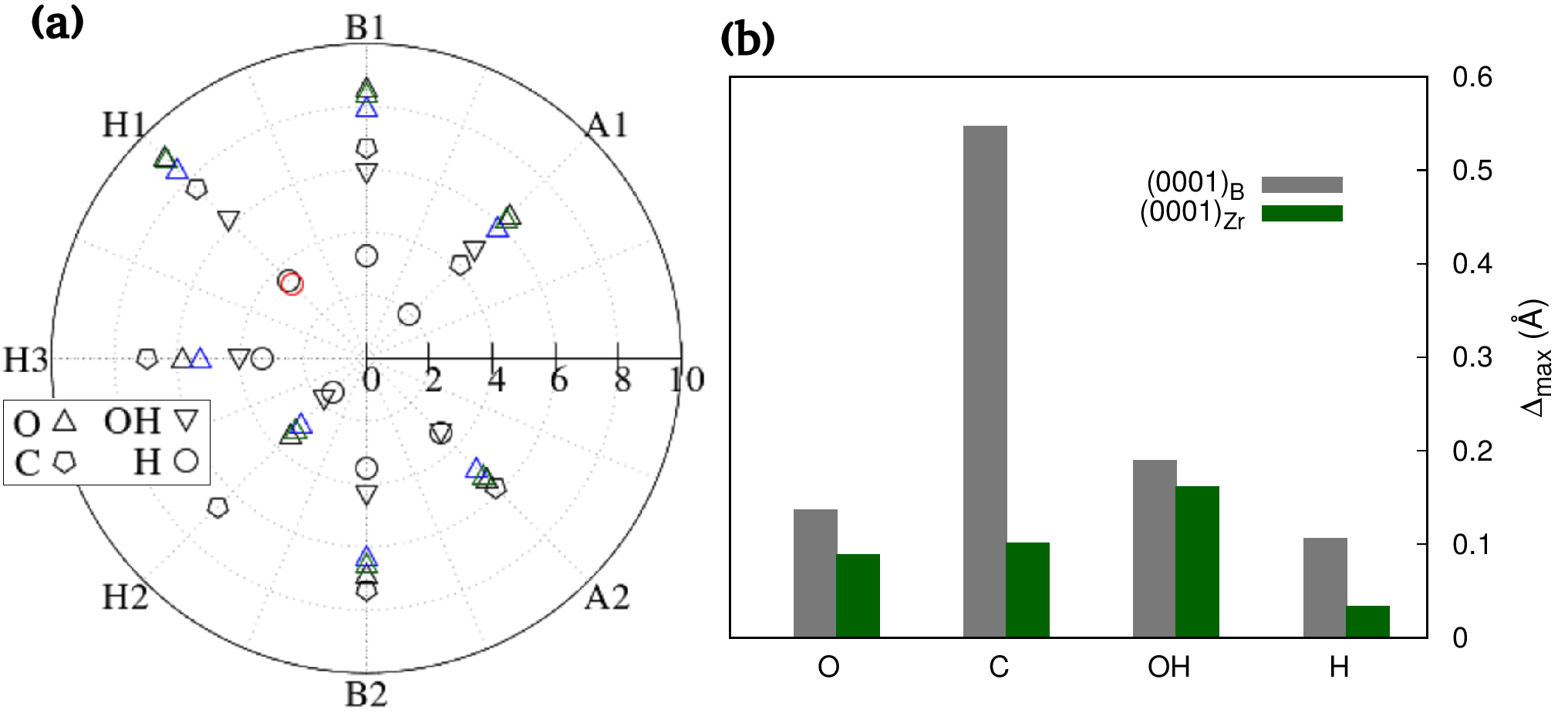}
\caption{(a) Polar diagram of the adsorption energies, $\Delta E_\mathrm{ads}$ (eV/atom), for the chemical 
species O, C, OH and H. Here H1, B1 and A1 (green shadow) are adsorption sites on (0001)$_\mathrm{Zr}$, 
and H2, H3, B2 and A2 (grey shadow) are on the (0001)$_\mathrm{B}$ surface. Our results in black are compared 
with those from Say\'{o}s's~\cite{Gamallo2013} in blue, Zhou's~\cite{Sun2016} in green and
Xiao's~\cite{ZENG201728} in red (note the latter two are shifted by half of the binding energies of O$_2$ and
H$_2$, respectively). (b) The disturbance of the adsorbent surface, $\Delta_\mathrm{max}$ (\AA), are measured by
the maximum displacement of the first atomic layer on the surface.}
\label{fig2}
\end{figure}

The fingerprint of chemisorption, namely a positive $\Delta E_\mathrm{ads}$ according to Eq.~(\ref{Etot}), is
observed for all the chemical species investigated, namely for O, C, H and OH, on both (0001)$_\mathrm{Zr}$ and
(0001)$_\mathrm{B}$. These are presented as a radial distribution in the polar diagram of Fig.~\ref{fig2}(a),
where the binding strength of a particular adsorbent grows with the distance from the center of the diagram and
the corresponding adsorption sites are marked along the rim. 

The rank of the relative adsorption strength for the different atomic species involved is O $\geq$ C $\geq$ OH
$\geq$ H on (0001)$_\mathrm{Zr}$ (O is the most strongly bonded species), which transforms into C $\geq$ O
$\geq$ OH $\geq$ H on (0001)$_\mathrm{B}$ (C is the most strongly bonded species). In more detail we find for O
the following $\Delta E_\mathrm{ads}$, 9.06, 3.44, 5.81, 8.59, 6.86, 6.43 and 5.42 (in eV/atom), respectively,
at the sites \textbf{H1}, \textbf{H2}, \textbf{H3}, \textbf{B1}, \textbf{B2}, \textbf{A1} and \textbf{A2}. These
values agree well with those from references~\cite{Gamallo2013,Sun2016}. Furthermore, $\Delta E_\mathrm{ads}$
for H is also in good agreement with that found by Xiao and coworkers~\cite{ZENG201728}. Note that the $\Delta E_\mathrm{ads}$ values extracted from Ref.~\cite{Sun2016} and Ref.~\cite{ZENG201728} and reported in Fig.~2(a)
are shifted by half of the binding energies of O$_2$ and H$_2$, respectively, due to the different reference
states chosen in such works. At the same time in the case of the Zr (0001) surface we can only compare with
results relative to metallic Zr~\cite{ZENG201728}, since those for ZrB$_2$ are not available in literature. To
the best of our knowledge $\Delta E_\mathrm{ads}$ information for C and OH on the (0001)$_\mathrm{Zr}$ and
(0001)$_\mathrm{B}$ surfaces of ZrB$_2$ are reported here for the first time.

We find the hollow and bond-bridge sites to be energetically more favourable than the atop ones. This is
generally true for both Zr- and B-terminated surfaces and for all the chemical species considered here.  In
fact, we compute $\Delta E_\mathrm{ads}$ on the hollow and bond-bridge sites of the (0001)$_\mathrm{Zr}$ surface
to range in the 3.22-9.06~eV/atom interval. These values are reduced to 1.52-7.39~eV/atom on the
(0001)$_\mathrm{B}$ surface. In contrast, the A1/A2 atop sites are significantly less preferred, presenting even
lower $\Delta E_\mathrm{ads}$ values. The tendency of O to adsorbe preferentially on hollow and bond-bridge
sites has been validated by experiments~\cite{HAYAMI1994} and other DFT-based calculations~\cite{Gamallo2013}.
Here we extend the same information to C and OH. 

Importantly, the chemical adsorption can induce a disturbance in the adsorbent surfaces. The maximum
displacement of the outmost surface layer resulting from the adsorption, $\Delta_\mathrm{max}$ (\AA), is
computed and shown in Fig.~\ref{fig2}(b). At variance with $\Delta E_\mathrm{ads}$, the (0001)$_\mathrm{B}$
surface atoms display $\Delta_\mathrm{max}$ values, for all the chemical species investigated, larger than those
of (0001)$_\mathrm{Zr}$. We find that on (0001)$_\mathrm{Zr}$, the OH adsorbate causes the strongest
disturbance, $\Delta_\mathrm{max}$ $\sim$ 0.16~\AA; while on (0001)$_\mathrm{B}$, it is the C adatom to give the
largest $\Delta_\mathrm{max}$ of 0.55~\AA. This fact can be explained by the light weight of B and the strong
chemical affinity between B and C (the existence of a stable B$_4$C compound with a melting point around
2800~K). In general, we find the (0001)$_\mathrm{B}$ surface to exhibit more prominent atomic disturbances than
the (0001)$_\mathrm{Zr}$ counterpart. 

In the following sections, we will discuss the adsorption mode of each molecule by examining the adsorption 
thermodynamics, the geometry of the adsorbate-adsorbent systems, the dissociation pathways and the associated 
activation energies. The adsorption and dissociation behaviour of the various molecules on B and Zr-terminated 
surfaces will also be compared.

\section{Adsorption and dissociation of O$_2$}
\subsection{Adsorption mechanism}
\begin{figure}[h!]
\centering
\includegraphics[scale=0.75]{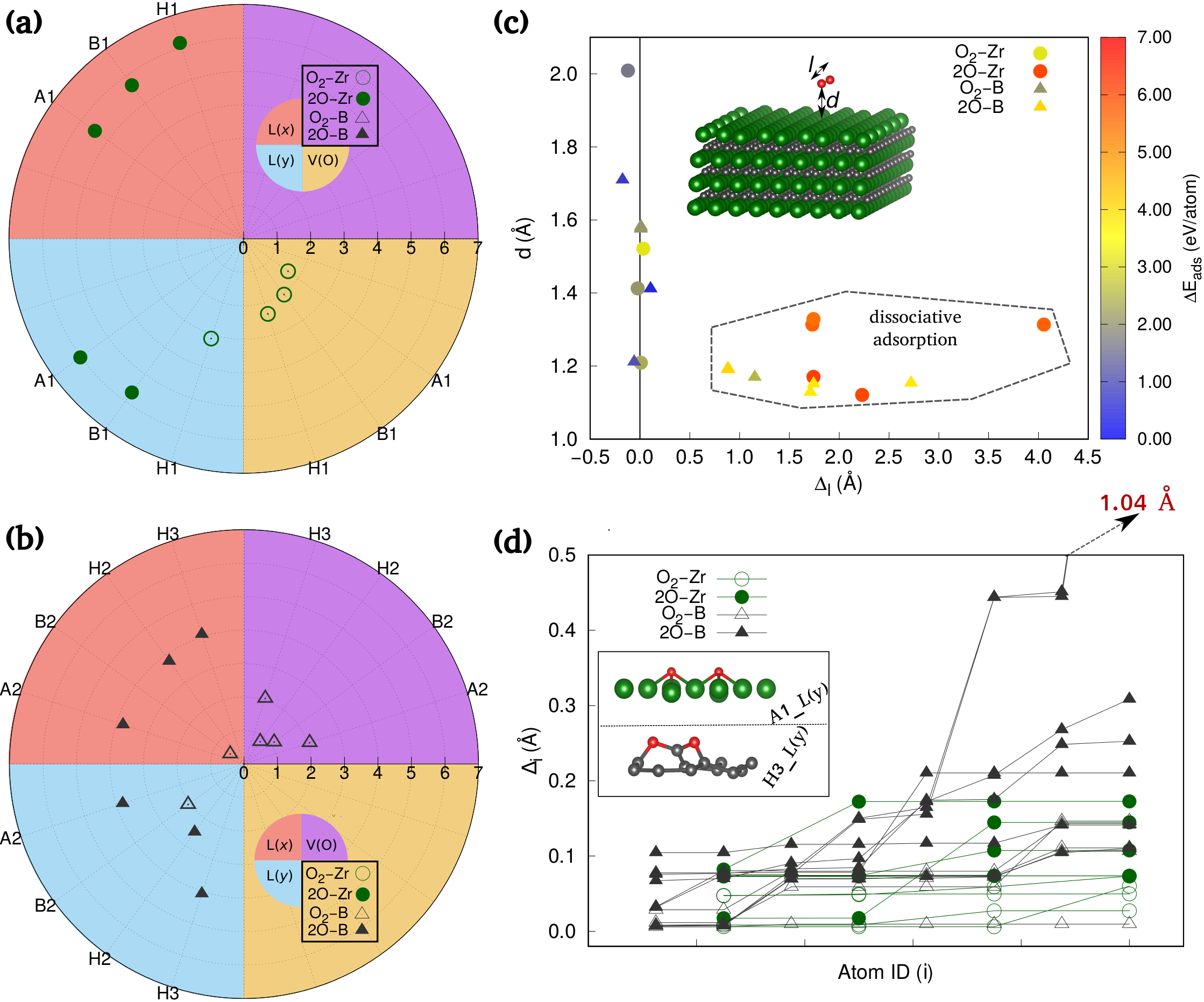}
\caption{Adsorption energies ($\Delta E_\mathrm{ads}$, eV/atom) for O$_2$ on the (a) (0001)$_\mathrm{Zr}$ and
(b) (0001)$_\mathrm{B}$ surfaces. Molecular (O$_2$) and  dissociative (2O) adsorptions are distinguished by
using open and filled symbols, respectively. The initial adsorption sites are marked along the rim of the polar
diagrams, while the approaching geometries of the O$_2$ molecule are encoded as background color. These are
further mapped as a function of the adsorption distance ($d$, \AA) and the change in the interatomic length
($\Delta_l$, \AA) in panel (c). The definitions of $d$ and $l$ are shown in the inset ($\Delta_l$ is the change
in $l$ due to the adsorption). The distortion of the adsorbent surfaces ($\Delta_i$) is plotted in panel (d) as
a function of the surface atoms index (see inset and main text). (Color code: Zr = green spheres; B: grey
spheres; O: red spheres).}
\label{fig3}
\end{figure}

As a general feature our results indicate that molecular O$_2$ is prone to dissociative adsorption when
laterally landing on the Zr- or B-terminated surfaces, but tends to adsorbe in the molecular form if vertically
approaching. These two cases are distinguished by the filled and the open symbols in the polar diagrams of
Fig.~\ref{fig3}(a) and (b), where we present the different adsorption energies of O$_2$ over
(0001)$_\mathrm{Zr}$ and (0001)$_\mathrm{B}$. The figure shows that all the dissociative adsorption starts from
lateral geometries (left halves of the polar diagrams), while most of the molecular adsorption events
originate from the vertical setups (right halves of the polar diagrams). Although dissociative adsorption of
O$_2$ has been observed in experiments~\cite{Aizawa2002a} in the past, the simple classification of the various
adsorption mechanisms based on the surface-approaching geometry is suggested here for the first time. 

As expected, the dissociated adsorption of O$_2$ is normally accompanied by a large variation of the O$_2$
geometry. In fact, the O-O interatomic length, $l$, of the dissociated O adatoms increases largely when compared
to the equilibrium bond length of 1.48~\AA, see $\Delta_l$ in Fig.~\ref{fig3}(c), indicating the O-O bond
breaking. The dissociated O adatoms then relax away from their initial adsorption location, and tend to finally
adsorb on the hollow (\textbf{H}) and bond-bridge (\textbf{B}) sites. This agrees well with the energetic
preference of the \textbf{H} and \textbf{B} sites discussed in section~\ref{AtomicAds}. Such behaviour has 
also been identified by Cheng et al. in Ref~\cite{Cheng2018}. 

At the same time, the dissociative adsorption of O$_2$ results in strongly bonded O adatoms on the
Zr-/B-surfaces, a feature clearly demonstrated by both the large adsorption energies, $\Delta E_\mathrm{ads}$,
and the short adsorption distances, $d$;  see Fig.~\ref{fig3}(c). In fact, $\Delta E_\mathrm{ads}$ for
dissociative adsorption on the Zr (B) surface is in the range 5.5-6.2 (2.5-4.1) eV/atom. These results can be
compared with with values of 6.57-7.08~eV/atom, reported for ultra-soft-pseudopotential calculations at the
GGA-PW91 level in Ref.~\cite{Cheng2018} for dissociative adsorption of O$_2$ on Zr surface. In contrast, 
molecular adsorption of O$_2$ on the Zr-terminated (B-terminated) surface has a much lower $\Delta E_\mathrm{ads}$ range, 1.6-3.1 (0.5-2.1)~eV/atom. The $d$ values distribute within the narrow range of 1.1-1.3
(1.1-1.2)~\AA~ for dissociated adatoms on Zr (B) surface, bond lengths that are considerably shorter than those
associated to molecular adsorbates (these can be as large as 2.0~\AA). The strong chemisorption of the O adatoms
on the ZrB$_2$ surfaces, with the corresponding large energy gain associated to the formation of the O-Zr and
O-B bonds, makes dissociative adsorption thermo-dynamically more favorable than molecular adsorption.

An important feature to note is that dissociative adsorption comes with sometimes severe surface distortions or
even surface reconstruction. In Fig.~\ref{fig3}(d) the displacement of each surface atoms are plotted against
their atom index. Dissociative adsorption of O$_2$ on (0001)$_\mathrm{Zr}$ activates surface reconstruction with
an atomic displacement of 0.17~\AA at the most. In contrast, dramatic effect is found for the \textbf{H3\_L(y)}
and \textbf{H3\_L(x)} on (0001)$_\mathrm{B}$ adsorbate-adsorbent systems, which present displacements as large
as 1.04~\AA. In these cases the two O adatoms position themselves as first nearest neighours on the bond-bridge
site after the O$_2$ dissociation. One B atom is then found to move above the original (0001)$_\mathrm{B}$
plane, thus breaking of the six-member ring symmetry of the (0001)$_\mathrm{B}$ surface; see the
\textbf{H3\_L(y)} illustration in the inset of Fig.~\ref{fig3}(d). These results suggest that the dissociative
adsorption of O$_2$ can trigger the reconstruction of the B-surface and therefore the easy formation of boron
oxides. Note that the melting temperature of B$_2$O$_3$, the most common among the boron oxides, is only 723~K.
This means that the formation of boron oxides during a high-temperature corrosion process will likely result in
significant B mass loss.

\subsection{Dissociation kinetics}
When an O$_2$ molecule approaches the surface in a lateral configuration dissociative adsorption takes place
without activation barrier, so that this is the most efficient reaction channel. However, in an O$_2$ gas one
cannot exclude situations where the molecule attacks ZrB$_2$ with a vertical geometry. For this reason we have
performed kinetic simulations for a number of candidate reactions involving both the vertical and lateral
channel. For each one of them, the reactants (precursors) acting as the initial states are chosen from the
adsorbed intermediates, while the dissociation products serving as the final states (FS) are selected among
dissociated adatoms. More specifically, we explore the following O$_2$ reaction paths (paths labelled with 
`a' are for reaction on the Zr-terminated surface, while those with `b' are for the B-terminated one):
\begin{equation}\nonumber
\begin{array}{l}
\left\{
\begin{array}{lllr}
\mathbf{O_2: H1\_V(O)} & \rightarrow & \mathbf{2O_{[ad, H1]} (1NN)} & \;\;\;\;\;\;\;\;\;\;\;\;\;\mathrm{(a1)} \\
\mathbf{2O_{[ad, H1]} (1NN)}  & \rightarrow &  \mathbf{2O_{[ad, H1]} (2NN)} & \;\;\;\;\;\;\;\;\;\;\;\;\;\mathrm{(a2)} \\
\end{array}
\right. \\
\\
\left\{
\begin{array}{lllr}
\mathbf{O_2: H2\_V(O)}  & \rightarrow & \mathbf{O_2: B2\_L(y)} & \;\;\;\;\;\;\;\;\;\;\;\;\;\;\;\mathrm{(b1)} \\
\mathbf{O_2: B2\_L(y)}  & \rightarrow & \mathbf{2O_{[ad, B2]} (1NN)} & \;\;\;\;\;\;\;\;\;\;\;\;\mathrm{(b2)} \\
\mathbf{2O_{[ad, B2]}(1NN)}  & \rightarrow & \mathbf{2O_{[ad, B2]} (2NN)} & \;\;\;\;\;\;\;\;\;\;\;\;\mathrm{(b3)} \\
\end{array}
\right.
\end{array}
\end{equation}

\begin{figure}[h!]
\centering
\includegraphics[scale=0.8]{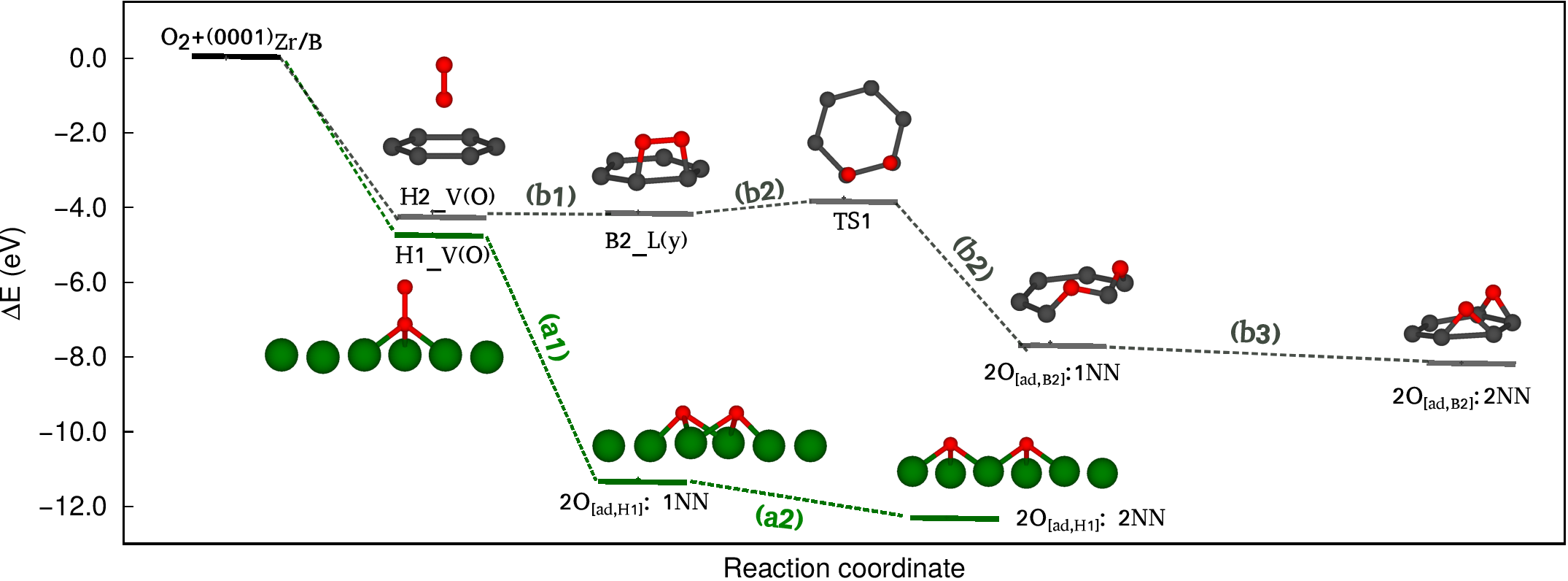}
\caption{Dissociation paths of O$_2$ on the (0001)$_\mathrm{Zr}$ surface via the (a1) and (a2) paths; and on the
(0001)$_\mathrm{B}$ one along the paths (b1), (b2) and (b3). In this second case one transition state (TS1),
with energy barrier 0.38~eV, is observed. Colour code: Zr = green; B grey; O = red.}
\label{fig4}
\end{figure}

When O$_2$ approaches the (0001)$_\mathrm{Zr}$ surface with the vertical geometry \textbf{H1\_V(O)}, it
dissociates into two O adatoms after having tilted so to assume a lateral configuration [path (a1) of
Fig.~\ref{fig4}]. The so-produced O adatoms can occupy two adjacent hollow sites, thus forming a first nearest
neighbour (1NN) pair, or they can diffuse further away from each other along the surface to take a second
nearest-neighbour (2NN) position, as described by the (a2) path [see Fig.~\ref{fig4}]. Both these paths are
characterised by the absence of a kinetic barrier. However, if the O adatoms diffuse towards the atop sites 
(energetically less favorable than the hollow ones), an energy barrier of 0.59~eV is found. In 
Ref.~\cite{Cheng2018} an activation barrier of 0.16~eV was determined for the vertical channel starting from
the \textbf{A1} site. The dissociation path for vertical adsorption discussed here, namely the combination of
paths (a1) and (a2), appears, however more energetically and kinetically advantageous. 

Turning now our attention to the dissociation of O$_2$ on (0001)$_\mathrm{B}$, we note that this is weakly
activated. The vertical configuration \textbf{H2\_V(O)} can firstly rotate into the lateral \textbf{B2\_L(y)},
as illustrated by path (b1). Thereafter, the O$_2$ adsorbate dissociates into two O-adatoms located at the
\textbf{B2} sites. This second step, path (b2), comes with a low activation barrier of 0.38~eV, a value in
close agreement with the previously calculated one of 0.27~eV from Say{\'o}s et al.~\cite{Gamallo2013}. Then,
the two O adatoms can occupy either the 2NN or the 1NN \textbf{B2} sites, by moving without energy barrier
along path (b3). Notably, the 1NN \textbf{B2} adsorbate geometry promotes a strong distortion of the B surface
by breaking the six-member B ring. For this reason the 2NN-\textbf{B2} configuration is more energetically and
kinetically favorable. 

\section{Adsorption and dissociation of H$_2$O}
\subsection{Adsorption mechanism}
\begin{figure}[h!]
\centering
\includegraphics[scale=0.75]{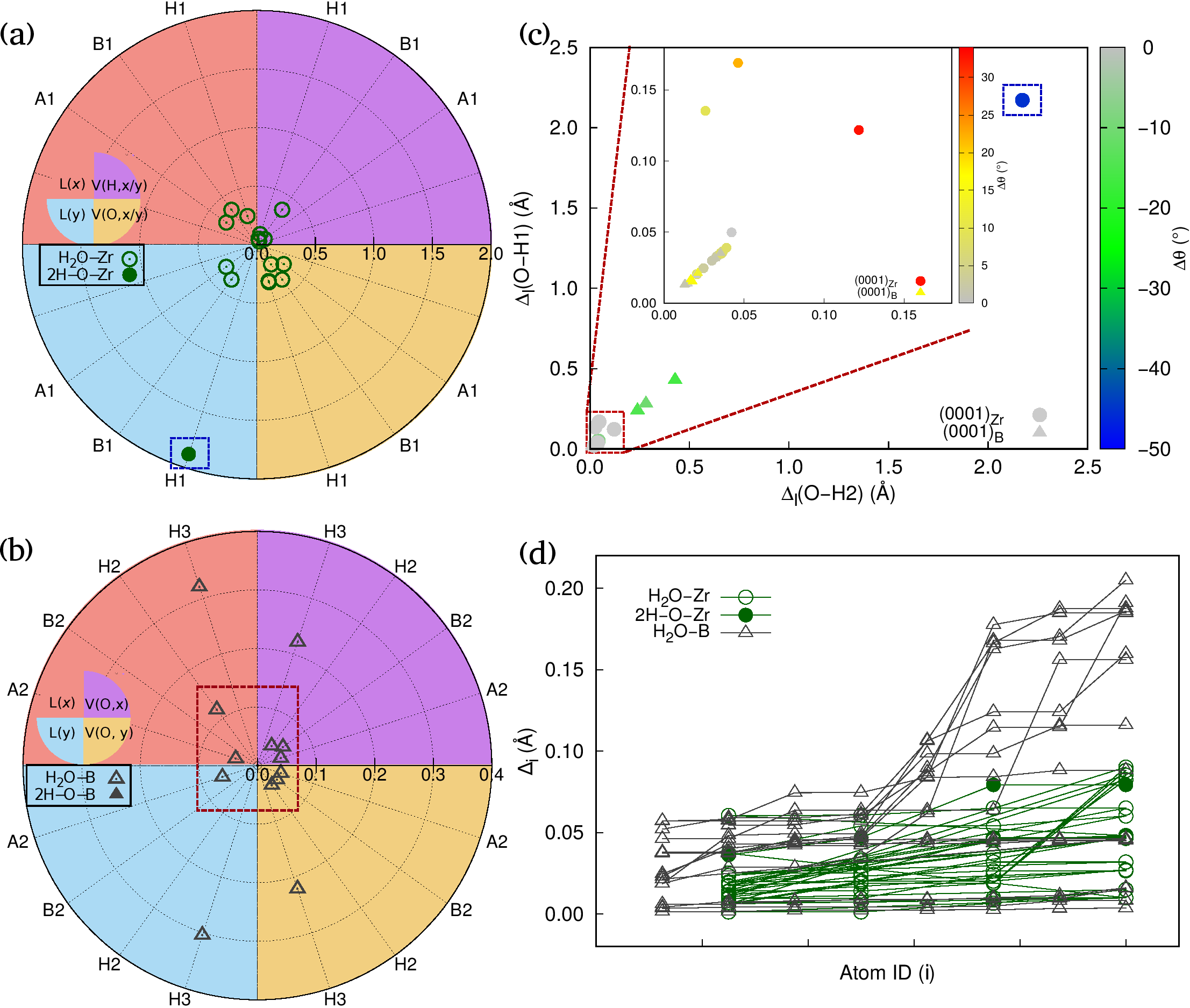}
\caption{Adsorption energies, $\Delta E_\mathrm{ads}$ (eV/atom), for H$_2$O on the (a) (0001)$_\mathrm{Zr}$ and
(b) (0001)$_\mathrm{B}$ surfaces. Molecular (H$_2$O) and dissociative (2H-O) adsorptions are distinguished by
using open and filled symbols, respectively.  The adsorption sites are marked along the rim of the polar
diagrams, while the approaching configurations of the H$_2$O molecules are encoded as background color. The
distortion of the H$_2$O geometry is analysed in panel (c) where the deviation of one H-O bond length,
$\Delta_l$(O-H1), from its equilibrium value is plotted against the other, $\Delta_l$(O-H2), and the color
scale maps the change in bond angle, $\Delta_\theta$. The inset zooms in the mild-distortion region. The
distortion of the adsorbent surfaces, $\Delta_i$, is plotted in panel (d) as a function of the surface atoms
index (see main text).}
\label{fig5}
\end{figure}

In general we find that water exhibits the characteristics of chemisorption on the (0001)$_\mathrm{Zr}$
surface, while physisorption dominates for the (0001)$_\mathrm{B}$ one. 
In more detail, when a H$_2$O molecule approaches the (0001)$_\mathrm{Zr}$ surface in a vertical position with
the O atom closer to the surface, then the typical $\Delta E_\mathrm{ads}$'s are between 207~meV/atom and
368~meV/atom, as illustrated in Fig.~\ref{fig5}(a). These values reduce by approximately a factor of four in
situations where it is the hydrogen side of the molecule to form contact to the surface, and the $\Delta E_\mathrm{ads}$ range is now 34-94~meV/atom. The only exception to this trend is for the \textbf{B1\_V(H, x)} 
geometry for which we compute a $\Delta E_\mathrm{ads}$ of 368~meV/atom. It must be noted that \textbf{B1\_V(H, x)}  is a highly symmetric structure, with the two protons taking the \textbf{H1} sites and the oxygen sitting
on the \textbf{B1} site.

In contrast, if the H$_2$O molecule is adsorbed on (0001)$_\mathrm{B}$, we will notice that both lateral and
vertical configurations result in weakly bonded absorbates, with the $\Delta E_\mathrm{ads}$'s clustering
around the 39-118~meV/atom interval. Only four configurations are observed with relatively large adsorption
energies, all above 200~meV/atom. These are all initiated with the O atom placed over the \textbf{H3} site, but
they consistently relax towards the \textbf{A2} one, a position that offers a more balanced adsorption
configuration for both O and H. 

Then we move to discuss the possibility of dissociative adsorption. This is observed on the
(0001)$_\mathrm{Zr}$ surface, although for significantly less configurations and with a much lower energy gain
than what found for O$_2$. In fact, only the \textbf{H1\_L(y)} configuration presents a clear evidence of
dissociative adsorption with a $\Delta E_\mathrm{ads}$ of 1.88~eV/atom; see Fig~\ref{fig5}(a). In fact,
in this case the final relaxed configuration shows an increase in the O-H bond length of more than 2.2~\AA,
while the bond angle decreases by about 50\degree~[see the blue symbol in Fig.~\ref{fig5}(c)]. No other
approaching geometry of H$_2$O on (0001)$_\mathrm{Zr}$ displays significant signs of molecular configuration 
change, with the bond lengths and bond angle remaining close to that of H$_2$O in the gas phase [see grey
symbols in the inset of Fig.~\ref{fig5}(c)]. A somewhat intermediate situation is found for H$_2$O attacking
the (0001)$_\mathrm{B}$ surface. In this case, three lateral configurations, namely \textbf{H2\_L(y)},
\textbf{B2\_L(x)} and \textbf{B2\_L(y)}, display significant bond-length expansion (0.2-0.5~\AA) and bond angle
contraction [green symbols in Fig.~\ref{fig5}(c)]. These distortions, however, are not large enough to define
the O-H bond breaking.

The disturbance of the surface atoms caused by the adsorption of H$_2$O is found to be minor when compared to
O$_2$ adsorption [see Fig.~\ref{fig5}(d)]. In fact, we find that the maximum displacement of the surface atoms
is always less that 0.21~\AA, while it was 1.04~\AA~for O$_2$. This is consistent with the $\Delta E_\mathrm{ads}$'s for water being significantly smaller (in absolute value) than that of O$_2$. To summarise,
our results indicate that H$_2$O adsorption is less active than O$_2$ adsorption, it is characterised by modest
adsorption energies and small geometry distortions of both the molecule and the surface.

\subsection{Dissociation kinetics}
Given the more complex structure of H$_2$O with respect to that of O$_2$, its surface dissociation kinetic is
also more complex. Here we consider the following potential reaction paths, where again `a'-labelled paths are
for (0001)$_\mathrm{Zr}$, while `b'-labelled ones are for (0001)$_\mathrm{B}$.
\begin{equation}\nonumber
\begin{array}{l}
\left\{
\begin{array}{lllr}
\mathbf{H_2O: B1\_V(O)} & \rightarrow & \mathbf{OH_{[ad, H1]} + H_{[ad,H1]}} & \;\;\;\;\;\;\;\;\;\;\;\;\mathrm{(a1)} \\
\mathbf{OH_{[ad, H1]} + H_{[ad,H1]}}  & \rightarrow &  \mathbf{2H_{[ad, H1]} + O_{[ad,H1]}} & \;\;\;\;\;\;\;\;\;\;\;\;\mathrm{(a2)} \\
\end{array}
\right. \\
\left\{
\begin{array}{lllr}
\;\mathbf{H_2O: B1\_V(O)}  & \;\;\;\;\;\;\rightarrow &  \mathbf{O_{[ad,B1]} + H_2(g)} & \;\:\;\;\;\;\;\;\;\;\;\;\;\;\;\;\;\;\;\mathrm{(a3)} \\
\end{array}
\right. \\
\left\{
\begin{array}{lllr}
\;\mathbf{H_2O: A1\_L(y)}  & \:\;\;\;\;\;\;\;\rightarrow &  \mathbf{2H_{[ad, H1]} + O_{[ad,H1]}} & \:\;\;\;\;\;\;\;\;\;\;\;\;\mathrm{(a4)} \\
\end{array}
\right. \\
\\
\left\{
\begin{array}{lllr}
\mathbf{H_2O: A2\_L(y)}  & \rightarrow & \mathbf{OH_{[ad, B2]} + H_{[ad, A2]}} & \;\;\;\;\;\;\;\;\;\;\;\;\;\;\mathrm{(b1)} \\
\mathbf{OH_{[ad, B2]}} + \mathbf{H_{[ad, A2]}}  & \rightarrow & \mathbf{2H_{[ad, A2]} + O_{[ad,B2]}} & \;\;\;\;\;\;\;\;\;\;\;\;\;\;\mathrm{(b2)} \\
\end{array}
\right. \\
\left\{
\begin{array}{lllr}
\;\mathbf{H_2O: A2\_L(y)}  & \;\;\;\;\;\;\;\rightarrow & \mathbf{2H_{[ad, B2]} + O_{[ad,B2]}} & \;\;\;\;\;\;\;\;\;\;\;\;\;\;\;\;\mathrm{(b3)} \\
\end{array}
\right.
\end{array}
\end{equation}

\begin{figure}[h!]
\centering
\includegraphics[scale=0.8]{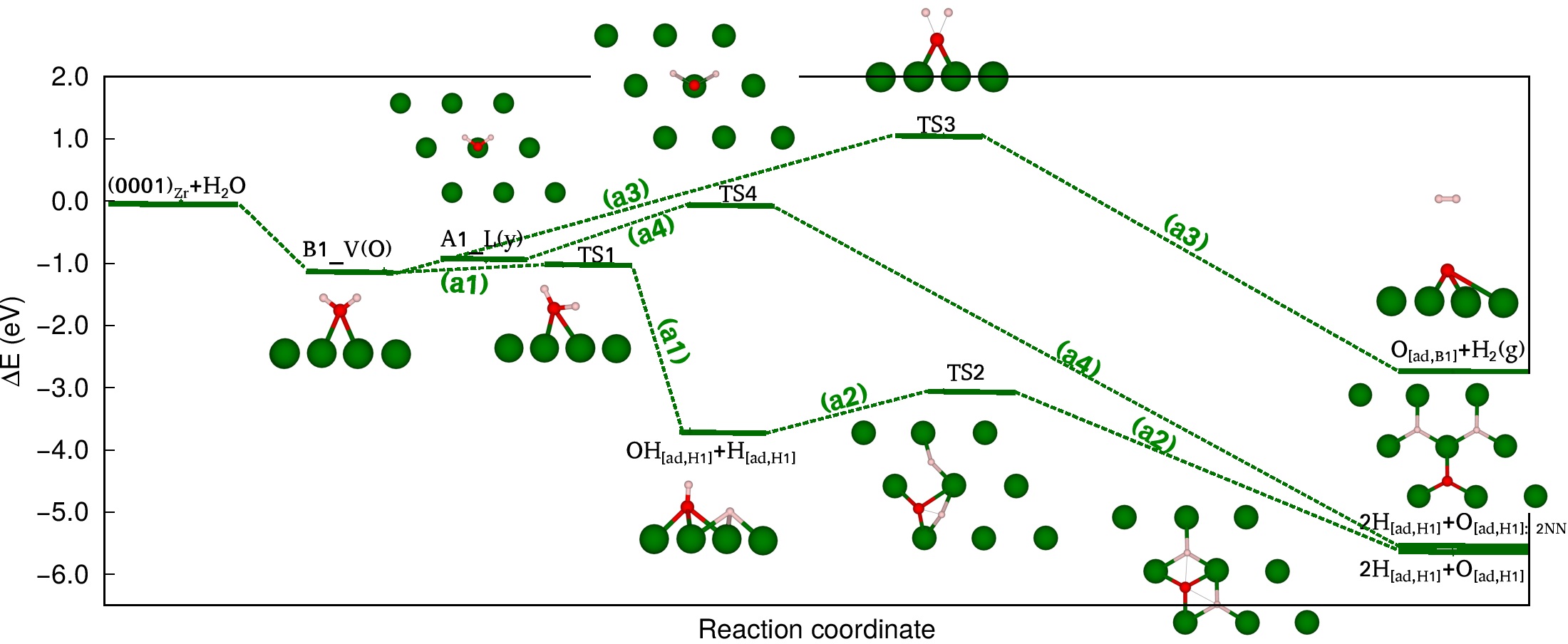}
\caption{Dissociation of H$_2$O on the (0001)$_\mathrm{Zr}$ surface. The initial states considered are
\textbf{B1\_{V(O)}} and \textbf{A1\_{L(y)}}. Along the (a1)+(a2) path the reaction proceeds by first producing
an OH group and a H atom and then by dissociating OH, so that the final state has two H and one O at adjacent
hollow sites. Along path (a3) a H$_2$O molecule approaching as \textbf{B1\_{V(O)}} releases an H$_2$ molecule,
leaving an O atom on the surface. The vertical path (a4) also exhibits direct dissociation into two H and one O
adatom, but these are placed at 2$^{nd}$ nearest neighbors positions. Four transition states are determined
with the following energy barriers, TS1: 0.09 eV, TS2: 0.65 eV, TS3: 2.16 eV, TS4: 0.94 eV. Colour code: Zr =
green; O = red; H = pink.}
\label{fig6}
\end{figure}

On (0001)$_\mathrm{Zr}$, we primarily explore the dissociation of H$_2$O starting from \textbf{B1\_{V(O)}}, an
energetically preferred adsorbate geometry. Our results are illustrated in Fig.~\ref{fig6}. When the molecule
moves along the (a1) path, it tilts and rotates so to first lose one proton, thereby forming an intermediate
state with the OH and H groups absorbed at two adjacent hollow sites. This intermediate state can further
dissociate into a final one, in which H and O adatoms all sit at hollow sites. We denote this second channel as
(a2). Two transition states, named TS1 and TS2, are detected for such combined (a1)+(a2) dissociation path,
with energy barriers, $E_\mathrm{b}$, of 0.09~eV and 0.64~eV, respectively. In an alternative path, (a3), a
H$_2$O molecule approaching with the \textbf{B1\_{V(O)}} geometry directly dissociates into atomic O by
releasing the two protons in the form of a H$_2$ gas molecule. The associated transition state, before H$_2$
release, has two protons separated by a distance of 1.09~\AA, while the O atom is adsorbed on the surface at 
a distance of about 1.56~\AA. The activation barrier for such path is, however, very high, 2.16~eV, making such
a path available only at extremely elevated temperatures. Finally, a lateral channel initiated at the
\textbf{A1$\_${L(y)}} structure can be activated by the expansion of the H-O-H bonds along path (a4). This is
characterised by an energy barrier, at the TS4 transition state, of 0.94~eV. Thus, we conclude that on
(0001)$_\mathrm{Zr}$ the (a1) path, with the low activation energy of 0.09~eV, is the most kinetically
favourable and dominates the dissociation of H$_2$O into the OH and H groups. Notably, this is the same 
dissociation reaction experimentally observed for H$_2$O on ZrC(001)~\cite{KITAOKA200123}. 

\begin{figure}[h!]
\centering
\includegraphics[scale=0.8]{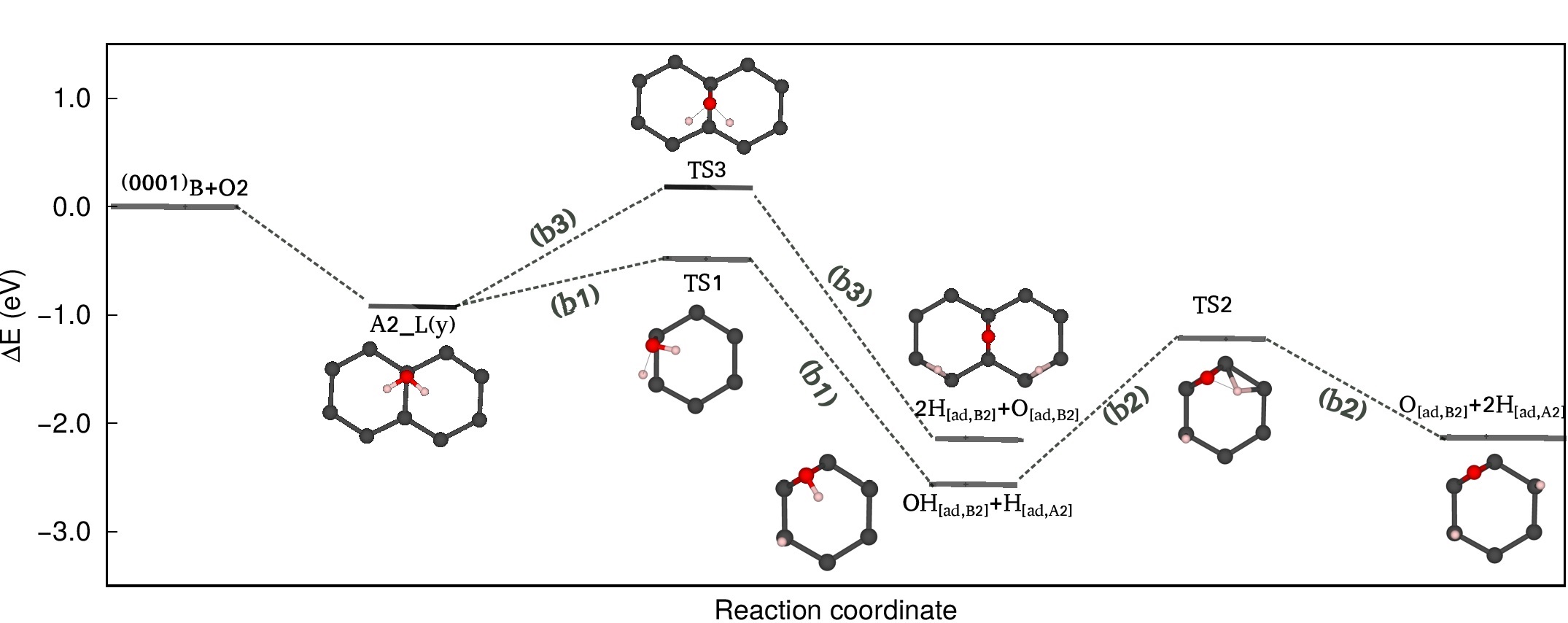}
\caption{Dissociation of H$_2$O on the (0001)$_\mathrm{B}$ surface initiated at the \textbf{A2\_{L(y)}}
geometry. We identify two main reaction channels, one involving sequential single deprotonation, (b1)+(b2), and
one associated to a double deprotonation, (b3). Three transition states are identified along these paths with
the energy barriers, TS1: 0.44 eV, TS2: 0.91 eV, TS3: 1.10 eV. Colour code: B = grey; O = red; H = pink.}
\label{fig7}
\end{figure}

Moving now to the (0001)$_\mathrm{B}$ surface, the dissociation of H$_2$O is studied starting from the stable 
adsorbate geometry \textbf{A2\_{L(y)}} along the two reaction paths illustrated in Fig.~\ref{fig7} [see also
reaction path list]. When going along (b1)+(b2) there is a first dissociation into an OH group and a H atom,
followed by the the decomposition of OH. The final dissociated geometry has O on the bond-bridge site and the
two H on the atop ones. For this reaction there are two activation barriers of 0.44~eV and 0.91~eV,
corresponding to a single deprotonation of H$_2$O and OH, respectively. The relatively high barrier for OH
deprotonation, together with the fact that the energy of fully dissociated configuration is higher than the one
in which the OH group remains intact on the surface, make this decomposition channel unlikely. In an 
alternative path, (b3), one O and two H adatoms are released simultaneously, ending up at three \textbf{B2}
sites. A transition state, TS3, with an energy barrier of 1.10~eV is identified along this path. TS3 is
characterised by the water molecule placed with the O atom at the bond-bridge site and the two H ones moving
toward the adjacent bridge positions (see Fig.~\ref{fig7}) with elongated O-H bonds. As such, also for the
(0001)$_\mathrm{B}$ surface we conclude that the dissociation of H$_2$O is likely to proceed by sequential
single deprotonations.

\section{Adsorption and dissociation of CO}
\subsection{Adsorption mechanism}
\begin{figure}[h!]
\centering
\includegraphics[scale=0.75]{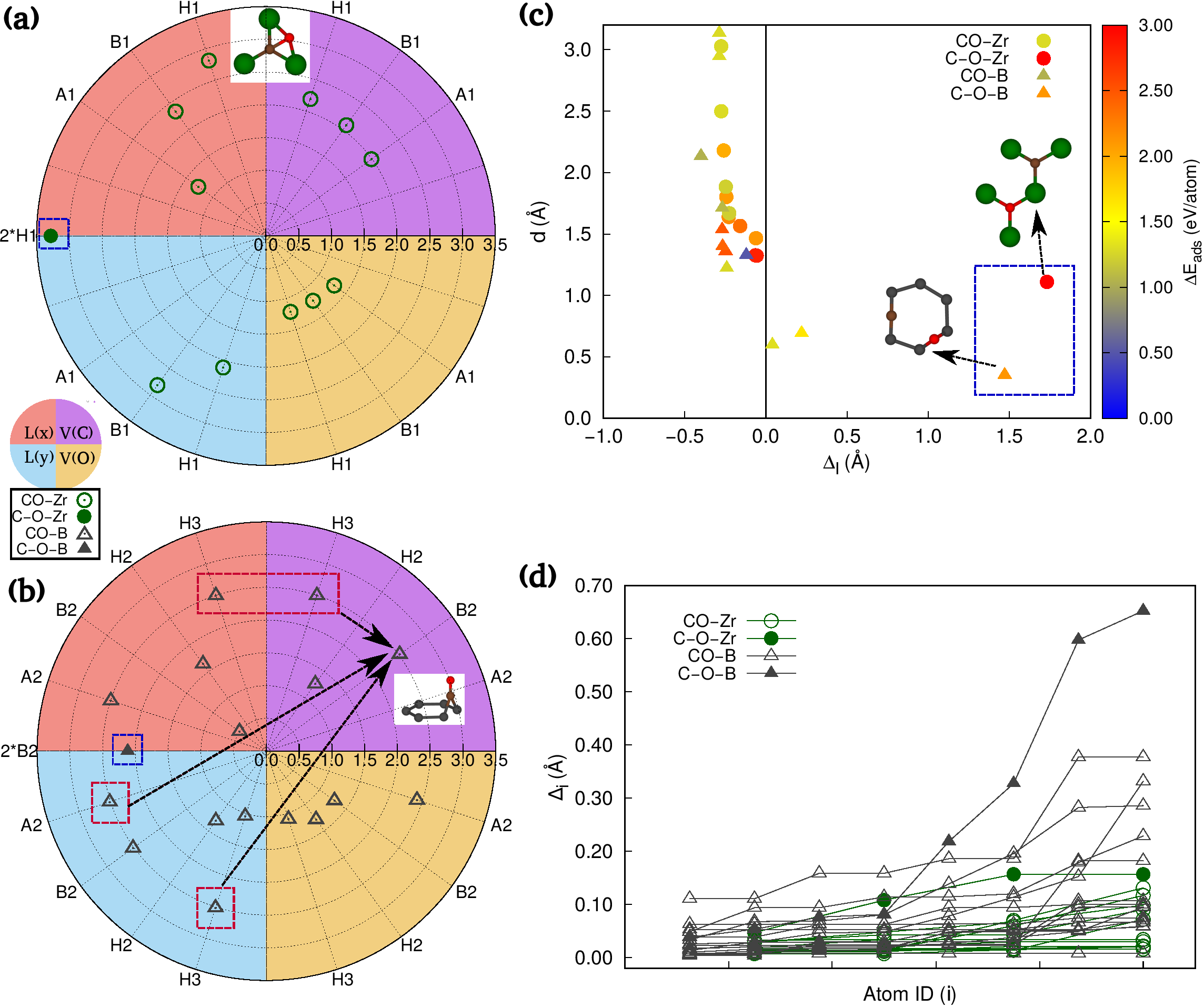}
\caption{Adsorption energies, $\Delta E_\mathrm{ads}$ (eV/atom), for CO on the (a) (0001)$_\mathrm{Zr}$ and (b)
(0001)$_\mathrm{B}$ surfaces. Molecular (CO) and dissociative (C + O) adsorptions are distinguished by using
open and filled symbols, respectively. The adsorption sites are marked along the rim of the polar diagrams,
while the approaching configurations of the CO molecules are encoded as background color. The insets illustrate
the preferred final configurations of the adsorbates. Note that on (0001)$_\mathrm{B}$, \textbf{H3\_L(x)},
\textbf{H3\_L(y)} and \textbf{A2\_L(y)} all transform into \textbf{B2\_V(C)}, so that their corresponding
$\Delta E_\mathrm{ads}$ are the same (red dashed boxes). In panel (c) we map $\Delta E_\mathrm{ads}$ as a
function of the adsorption distance, $d$ (\AA), and the change in interatomic lengths, $\Delta_l$ (\AA). 
The distortion of the adsorbent surfaces, $\Delta_i$ (\AA), is plotted in panel (d) as a function of the
surface atoms index (see main text). Colour code: Zr = green; B = grey; C = brown; O = red.}
\label{fig8}
\end{figure}

In this section we investigate the energetics of CO adsorption. In general, our calculations show that CO can
be chemisorbed on both ZrB$_2$ surfaces with typical adsorption strengths falling in between those of O$_2$ and
H$_2$O. In fact, we find $\Delta E_\mathrm{ads}$'s on the Zr-terminated (B-terminated) surface within the range
1.22-2.81~eV/atom (0.51-2.51~eV/atom), see Figs.~\ref{fig8}(a) and \ref{fig8}(b). 

On (0001)$_\mathrm{Zr}$ the lateral CO configurations are in general more energetically stable than the
vertical ones, similarly to what found for O$_2$ and H$_2$O adsorption. It is worth mentioning that the initial
geometries \textbf{H1\_L(x)} and \textbf{B1\_L(y)} can rotate into tilted structures and finally settle with O
on the \textbf{B1} and C on the \textbf{H1} site. The top view of the final configuration of \textbf{H1\_L(x)}
is illustrated in the inset of Fig.~\ref{fig8}(a). Note that the \textbf{B1} and \textbf{H1} adsorption sites
are energetically rather similar, therefore in competition for adsorption. 

In contrast, on (0001)$_\mathrm{B}$ it is the vertical adsorption to be more energetically favourable. Furthermore, we find that when CO is landing on the surface in a lateral configuration it can always rotate to
assume a vertical structure during the geometry relaxation. For example, we find that \textbf{H3\_L(x)},
\textbf{H3\_L(y)} and \textbf{A2\_L(y)} all transform into \textbf{B2\_V(C)}. As such their adsorption
energies, $\Delta E_\mathrm{ads}$ turn out to be all the same, as highlighted by the red dashed boxes of
Fig.~\ref{fig8}(b), while their final adsorbate geometry is shown in the inset. In the same way, we find that
the lateral configurations \textbf{H2\_L(x)} and \textbf{H2\_L(y)} all relax into the tilted \textbf{H2\_V(C)}
one. We believe that such strong preference for vertical geometries is associated to the chemical affinity
difference between C-B and O-B. 

Importantly, for CO we do not find any evidence of direct dissociative adsorption on either (0001)$_\mathrm{Zr}$ or (0001)$_\mathrm{B}$, a fact that holds true for both lateral and vertical approaching
geometries. The relative small variations in bond length of the various configurations are presented in
Fig.~\ref{fig8}(c). In order to complete our adsorption energy study we have also specifically constructed two
dissociated configurations, namely \textbf{O$_\mathbf{[ad, H1]}$-C$_\mathbf{[ad,H1]}$} and
\textbf{O$_\mathbf{[ad, B2]}$-C$_\mathbf{[ad, B2]}$}, whose
geometries are illustrated in the insets of Fig.~\ref{fig8}(c). These present $\Delta E_\mathrm{ads}$ values
rather close to those found for molecular adsorbates, as it can be noted by comparing the empty and filledsymbols in Figs.~\ref{fig8}(a) and \ref{fig8}(b). Such close energetics between molecular and dissociated
adsorbates is indicative of a weak thermodynamic driving force to break the C-O bond during the surface
adsorption.

Finally, the atomic distortions of the surface atoms due to the adsorption of CO are displayed in
Fig.~\ref{fig8}(d) as a function of the surface atom index. For molecular adsorption we find that B atoms move
within the interval 0.004-0.38~\AA, while the range for Zr is significantly more narrow, 0.008-0.13~\AA. Large
distortions are found only for dissociated configurations on the B-terminated surface, 0.65~\AA, while they
remain modest on the Zr-terminated one, 0.16~\AA. Thus, we conclude that the surface distortions caused by CO
is much less severe than that associated to O$_2$, but more pronounced than that of H$_2$O. This is consistent
with the same ranking found for the adsorption energies. As such, there is compelling evidence that O$_2$ is
the most active chemical molecule on ZrB$_2$ surfaces, followed by CO and lastly by H$_2$O. 
\subsection{Dissociation kinetics}
Since molecular CO can be adsorbed both laterally and vertically on (0001)$_\mathrm{Zr}$, and only vertically on
(0001)$_\mathrm{B}$, we decide to examine the following dissociation paths,
\begin{equation}\nonumber
\begin{array}{l}
\left\{
\begin{array}{lllr}
\mathbf{CO: H1\_V(C)} & \rightarrow & \mathbf{CO: B1\_L(y)} & \:\;\:\;\;\;\;\:\;\;\;\;\:\;\;\;\;\:\;\;\;\;\;\;\;\;\;\;\;\;\;\;\;\mathrm{(a1)} \\
\mathbf{CO: B1\_L(y)}  & \rightarrow &  \mathbf{C_{[ad, H1]} + O_{[ad,H1]}} & \:\;\:\;\;\;\;\:\;\;\;\;\:\;\;\;\;\:\;\;\;\;\;\;\;\;\;\;\;\;\;\;\;\mathrm{(a2)} \\
\end{array}
\right. \\
\left\{
\begin{array}{lllr}
\;\mathbf{CO: B2\_V(C)}  & \rightarrow &  \mathbf{C_{[ad,B2]} + O_{[ad,B2]} (2NN)} & \;\;\;\:\;\;\;\;\;\;\;\;\;\;\;\;\;\;\;\;\;\mathrm{(b1)} \\
\end{array}
\right. 
\end{array}
\end{equation}

\begin{figure}[h!]
\centering
\includegraphics[scale=0.8]{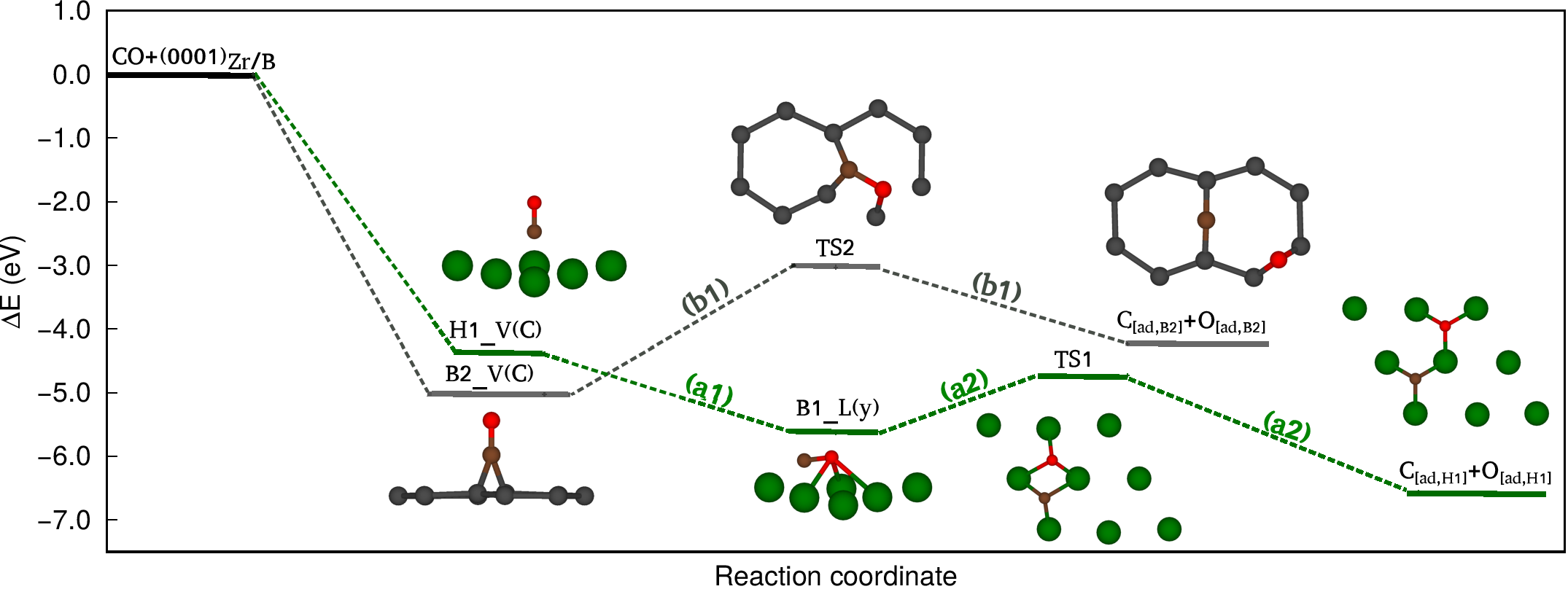}
\caption{Dissociation of CO over (0001)$_\mathrm{Zr}$ (green line) and (0001)$_\mathrm{B}$ (grey line) along
the paths (a1)+(a2) and (b1), respectively. Two transition states, TS1 and TS2, are identified respectively for
(0001)$_\mathrm{Zr}$ and (0001)$_\mathrm{B}$ with energy barriers 0.90~eV and 2.01~eV. Colour code: Zr = green;
B = grey; C = brown; O = red.}
\label{fig9}
\end{figure}

On (0001)$_\mathrm{Zr}$ we start by taking \textbf{H1\_V(C)} as initial CO geometry. This is thermodynamically
less stable than the lateral configuration, \textbf{B1\_L(y)}, and in fact our NEB calculation reveals that
\textbf{H1\_V(C)} can transform into \textbf{B1\_L(y)} without any activation barrier [see path (a1) in
Fig.~\ref{fig9}]. Then, the lateral adsorbate \textbf{B1\_L(y)} can decompose into two final adatoms,
\textbf{C$_\mathbf{[ad,H1]}$} and  \textbf{O$_\mathbf{[ad,H1]}$}, located at 2$^\mathrm{nd}$ nearest-neighbour
positions. The decomposition takes place along the (a2) path of Fig.~\ref{fig9}. Along such reaction trajectory
one encounters a transition state, TS1, where the two adatoms are in a 1$^\mathrm{st}$ nearest-neighbour
position. The transition state corresponds to the saddle point of the O migration and the associated activation
barrier is 0.90~eV. 

In contrast, on (0001)$_\mathrm{B}$ the preferred adsorption geometry is vertical. In fact, any lateral
configuration investigated rotates to a vertical position upon DFT relaxation, namely it cannot be stabilised
even as metastable state. As representative configuration we then take \textbf{B2\_V(C)}, which possesses the
lowest adsorption energy on (0001)$_\mathrm{B}$ [see (b1) channel in Fig.~\ref{fig9}]. The final state
considered is made of two adatoms, \textbf{C$_\mathbf{[ad,B2]}$} and \textbf{O$_\mathbf{[ad,B2]}$}, occupying
2$^\mathrm{nd}$ nearest-neighbour positions. Such reaction path contains a transition state, TS2, in which the
B hexagonal structure is broken and the CO molecule effectively replaces on B atom. The activation barrier at
the transition state is rather large, namely 2.01~eV. 

As noted before, the chemical activity of CO is located in between that of O$_2$ (the most reactive species)
and that of H$_2$O (the less reactive species), and it can result in surface reconstruction, similarly to the
O$_2$ case. However, the dissociation kinetics of CO shows a behavior totally different from that of O$_2$. In
fact bond breaking in O$_2$ can take place without kinetic activation, while that of CO demands an activation
energy of 0.90~eV on (0001)$_\mathrm{Zr}$ and 2.01~eV on (0001)$_\mathrm{B}$. Therefore, we conclude that
ZrB$_2$ oxidation can dominate at relatively low temperatures, while corrosion caused by CO requires elevated
temperatures to be active. 

\section{Discussion}

ZrB$_2$, a strongly bonded compound presenting alternating Zr and B atomic planes, is likely to present a grain
structure dominated by (0001)$_\mathrm{Zr}$ and (0001)$_\mathrm{B}$ facets, which are also the most
energetically stable open surfaces under a broad range of growth conditions. These surfaces undergo corrosive
attack in the aggressive atmosphere, containing O$_2$, H$_2$O and CO, typical of real-life combustion in
aerospace applications. Here we investigate the first steps in the corrosion process, namely we focus on the
initial surface reactions involving molecular adsorption and dissociation. Our calculations confirm that
(0001)$_\mathrm{Zr}$ generally promotes strongly bound adsorption and weakly activated dissociation. Therefore,
the transition-metal-terminated (0001)$_\mathrm{Zr}$ surface is found to be more chemical active than its
(0001)$_\mathrm{B}$ counterpart. This is not a surprise given the high reactivity of early transition metals.

Turning our attention to the (0001)$_\mathrm{B}$ surface, this is characterised by a substantial reconstruction
upon both CO and O$_2$ dissociation. The limiting kinetic process in the first case is determined by a large
activation barrier of 2.01~eV. In contrast, O$_2$ dissociation requires a much more modest activation energy, 
0.38~eV, so that the formation of boron oxides is likely to be determined by the O$_2$ supply. The final
reaction product, B$_2$O$_3$, has a relatively low melting point ($\sim$700~K), while evaporation takes place
above 1500~K. Thus, we expect that at intermediate temperatures B layers can be easily depleted due to the
presence of O$_2$, a mechanism similar to the C loss observed in ZrC~\cite{Kato_2000}. This appears to be the
most efficient mechanism for B depletion during combustion.

Furthermore, also at the (0001)$_\mathrm{Zr}$ surface the dominant reaction is the dissociative adsorption of
O$_2$, which proceeds spontaneously without activation. Importantly, the dissociated O adatoms can easily
diffuse at the surface with migration paths presenting a barrier of 0.59~eV, but with the presence of
barrier-less migration trajectories as well. When these results are combined together we can unambiguously
suggest that the initial oxidation process of ZrB$_2$ can proceed rapidly. A possible strategy for limiting the
corrosion may then be that of inserting additives into the ZrB$_2$ matrix. Alternatively one can develop
appropriate surface treatments, such as pre-oxidation. This may form a protective ZrO$_2$ layer over ZrB$_2$. 
Importantly, ZrO$_2$ has high thermal stability (melting temperature, $T_m$ $\leq$ 2900~K), and the interface
attachment between ZrB$_2$ and ZrO$_2$ can be controlled by controlling the ZrO$_2$ polymorphism.

\section{Conclusion}

In this work, we have presented a thorough atomistic modelling of the adsorption and dissociation of O$_2$,
H$_2$O and CO over two of the most stable ZrB$_2$ surfaces. Several conclusions can be drawn from our results.
Firstly, all the chemical species investigated, namely O, C, OH and H, display strong chemisorption on ZrB$_2$,
with O and C showing the largest adsorption energies. In general, the preferred adsorption sites on both
Zr- and B-terminated surfaces are the hollow and bridge ones. Secondly, we have found that O$_2$ dissociative
adsorption prevails when the molecule laterally approaches the surface, while molecular adsorption dominates
for vertical geometries. The same trend is also found for the other molecules, with the exception of CO on
(0001)$_\mathrm{B}$. Thirdly, we have determined that the chemical activity of the various adsorbents follows
the ranking: O$_2$, CO and H$_2$O, with O$_2$ being the most active agent. H$_2$O is the less reactive of the
molecules studies and tends to be chemisorbed on (0001)$_\mathrm{Zr}$ and physisorbed on (0001)$_\mathrm{B}$.
The kinetic barriers for dissociation are instead ordered as O$_2$, H$_2$O and CO, with CO presenting
anactivation barrier of 0.90~eV on (0001)$_\mathrm{Zr}$ and of 2.01~eV on (0001)$_\mathrm{B}$. This makes O$_2$
the most aggressive chemical agent for ZrB$_2$. Finally, we have determined that the Zr-surface is chemically
more active than the B-terminated one, at least to the chemical agents studied here. Considering the low
melting point of the B-containing products, the B-surfaces are likely to be destroyed under O$_2$ and/or CO
attack, a mechanism likely to determine B mass lost during combustion.

As it stands, our work offers a complete view of the microscopic mechanism for reaction of O$_2$, CO and H$_2$O
on ZrB$_2$ surfaces. This is important to determine the first steps in the corrosion process. Future work on
the effects of the intermediate products and of the temperature is necessary for defining a complete picture of
oxidation and corrosion in UHTC materials. 

\section*{Acknowledgement}
This research was supported by the European Union's Horizon 2020 ``Research and innovation programme''
under the grant agreement No.685594 (C$^3$HARME). Computational resources have been provided by the Irish
Center for High-End Computing (ICHEC) and the Trinity Centre for High Performance Computing (TCHPC). Y.Z. 
would like to thank the support from the internal grant of Yanshan University and the 100 Talents Programme of 
Hebei province.
%It was also supported by the 100 Talents Plan of Hebei Province.

%\bibliography{library}
\section*{Reference}
\providecommand{\latin}[1]{#1}
\makeatletter
\providecommand{\doi}
  {\begingroup\let\do\@makeother\dospecials
  \catcode`\{=1 \catcode`\}=2 \doi@aux}
\providecommand{\doi@aux}[1]{\endgroup\texttt{#1}}
\makeatother
\providecommand*\mcitethebibliography{\thebibliography}
\csname @ifundefined\endcsname{endmcitethebibliography}
  {\let\endmcitethebibliography\endthebibliography}{}

\end{document}